\documentclass[twocolumn,twoside,slac]{revtex4}
\usepackage[dvips]{graphicx}
\usepackage{fancyhdr}
\begin{document}

\title{\raggedleft \normalsize GriPhyN Report 2003-16\\
CMS CR 2003-015\\
physics/0306008\\
\raggedright \Large Virtual Data in CMS Analysis}

%

\author{A.~Arbree, P.~Avery, D.~Bourilkov, R.~Cavanaugh, J.~Rodriguez}
\affiliation{University of Florida, Gainesville, FL 32611, USA}
\author{G.~Graham}
\affiliation{FNAL, Batavia, IL 60510, USA}
\author{M.~Wilde}
\affiliation{ANL, Argonne, IL 60439, USA}
\author{Y.~Zhao}
\affiliation{University of Chicago, Chicago, IL 60637, USA}
\author{ }
\affiliation{ }
\author{Presented by D.Bourilkov, bourilkov@mailaps.org}

\begin{abstract}
   The use of virtual data for enhancing the collaboration
between large groups of scientists is explored in several
ways:

$\bullet$  by defining ``virtual'' parameter spaces 
     which can be searched and shared in an organized way by a
     collaboration of scientists in the course of their analysis

$\bullet$  by providing a mechanism to log the provenance of results and the
     ability to trace them back to the various stages in the analysis
     of real or simulated data

$\bullet$  by creating ``check points'' in the course of an
     analysis to permit collaborators to
     explore their own analysis branches by refining
     selections, improving the signal to background ratio, varying the 
     estimation of parameters, etc.

$\bullet$  by facilitating the audit of an analysis and the reproduction of
     its results by a different group, or in a peer review context.

   We describe a prototype for the analysis of data from the CMS experiment
based on the virtual data system {\tt Chimera} and the object-oriented data
analysis framework {\tt ROOT}. The {\tt Chimera} system is used to chain
together several steps in the analysis process including
the Monte Carlo generation of data, the simulation of
detector response, the reconstruction of physics objects and their
subsequent analysis, histogramming and visualization using the {\tt ROOT}
framework.

\end{abstract}

\maketitle

\thispagestyle{fancy}


\section{INTRODUCTION}

A look-up in the Webster dictionary gives:

{\bf vir·tu·al}\\
Function: adjective\\
Etymology: Middle English, possessed of certain physical virtues,
from Medieval Latin virtualis, from Latin virtus strength, virtue.

In this contribution we explore the virtue of virtual data in the scientific
analysis process, taking as an example the coming generation of high energy
physics (HEP) experiments at the Large Hadron Collider (LHC),
under construction at CERN close to Geneva.

Most data in contemporary science are the product of increasingly complex
computations and procedures applied on the raw information coming from
detectors (the ``measurements'') or from
numeric simulations - e.g. reconstruction, calibration, selection,
noise reduction, filtering , estimation of parameters etc.
High energy physics and many other sciences are increasingly CPU and data
intensive. In fact, many new problems can only be addressed at the high
data volume frontier. In this context, not only data analysis transformations,
but also the detailed log of how those transformations were applied,
become a vital intellectual resource of the scientific community.
The collaborative processes of these ever-larger groups require
new approaches and tools enabling 
the efficient sharing of knowledge and data across a geographically
distributed and diverse environment.

The scientific analysis process demands the precise tracking of
how data products are to be derived, in order
to be able to create and/or recreate them on demand.
In this context virtual data are data products with a well defined method
of production or reproduction. The concept of ``virtuality'' with respect to
existence means
that we can define data products that may be produced in the
future, as well as record the ``history'' of products that exist now
or have existed at some point in the past.
Recording and discovering the relationships can be important for many
reasons - some of them, adapted from~\cite{chimera} to high energy physics
applications, are given below:
\begin{itemize}
\item ``I have found some interesting data, but I need to know exactly what
corrections were applied before I can trust it.''
\item ``I have detected a muon calibration error and want to know which
derived data products need to be recomputed.''
\item ``I want to search a huge database for rare electron events. If a
program that does this analysis exists, I will not have to reinvent the
wheel.''
\item ``I want to apply a forward jet analysis to 100M events. If the
results already exist, I will save weeks of computation.''
\end{itemize}
We need a ``virtual data management'' tool that can ``re-materialize'' data
products that were deleted, generate data products that were defined but
never created, regenerate data when data dependencies or algorithms change,
and/or create replicas at remote locations when recreation is more efficient
than data transfer.
\begin{figure*}[t]
\centering
\resizebox{0.95\textwidth}{0.45\textheight}{\includegraphics{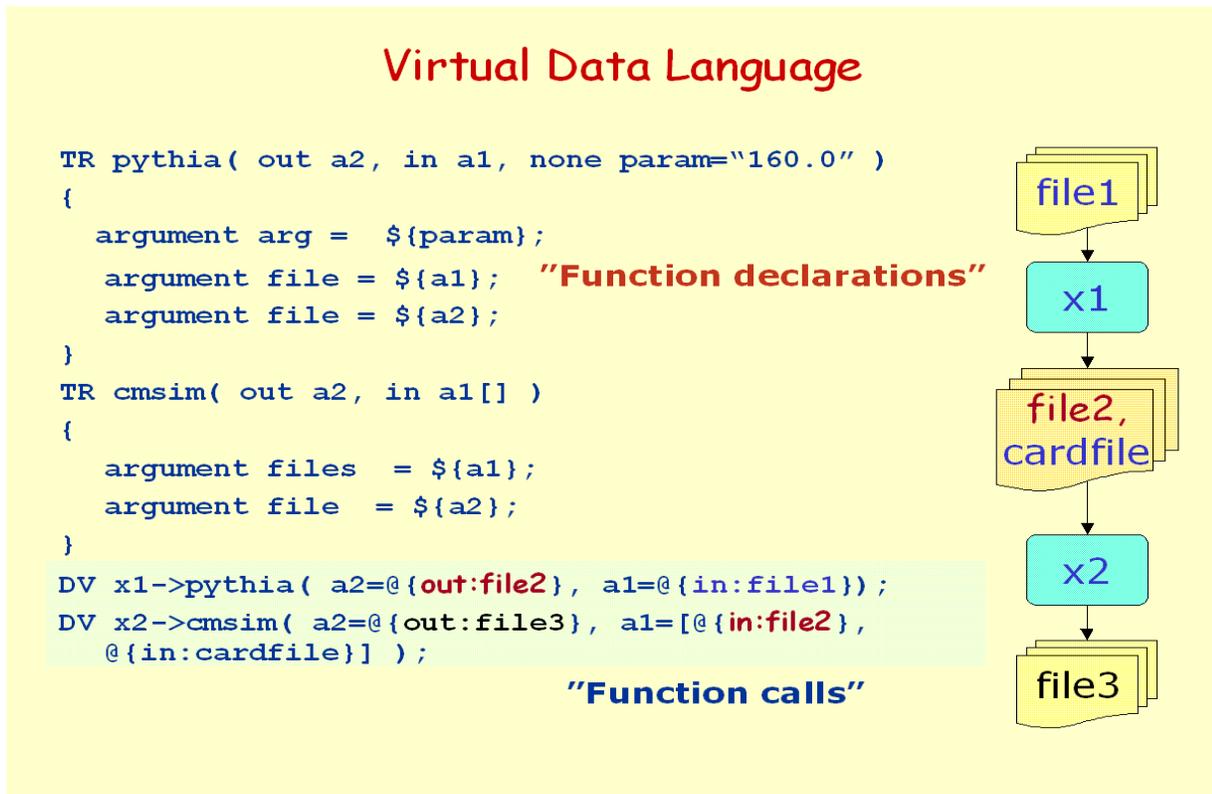}}
\caption{Example of a VDL description for a pipeline with two steps.} \label{dbchep03-07}
\end{figure*}

The virtual data paradigm records data provenance by tracking
how new data is derived from transformations on other data~\cite{chimera}.
It focuses on two central concepts: transformations and derivations.
A {\em transformation} is a computational procedure used to derive data.
A {\em derivation} is an invocation of such a procedure, resulting in
the instantiation of a potential data product.
{\em Data provenance} is the exact history of any existing (or virtual)
data product.
Often the data products are large datasets, and the management of dataset
transformations is critical to the scientific analysis process.

From the scientist's point of view, data trackability and result auditability
are crucial, as the reproducibility of results is fundamental to the nature of
science. To support this need we require and envision something like a
``virtual logbook'' that provides the following capabilities:
\begin{itemize}
\item easy sharing of tools and data to facilitate collaboration -
all data comes complete with a ``recipe'' on how to produce or reproduce it;
\item individuals can discover in a fast and well defined way other scientists'
work and build from it;
\item different teams can work in a modular, semi-autonomous fashion; they
can reuse previous data/code/results or entire analysis chains;
\item the often tedious procedures of repair and correction of data can
be automated using a paradigm similar to that which "make" implements for
rebuilding application code;
\item on a higher level, systems can be designed for workflow management and
performance optimization, including the tedious processes of staging in data
from a remote site or recreating it locally on demand (transparency with
respect to location and existence of the data);
\end{itemize}

\section{CHIMERA - THE GRIPHYN VIRTUAL DATA SYSTEM}

\begin{figure*}[t]
\centering
\resizebox{0.95\textwidth}{0.45\textheight}{\includegraphics{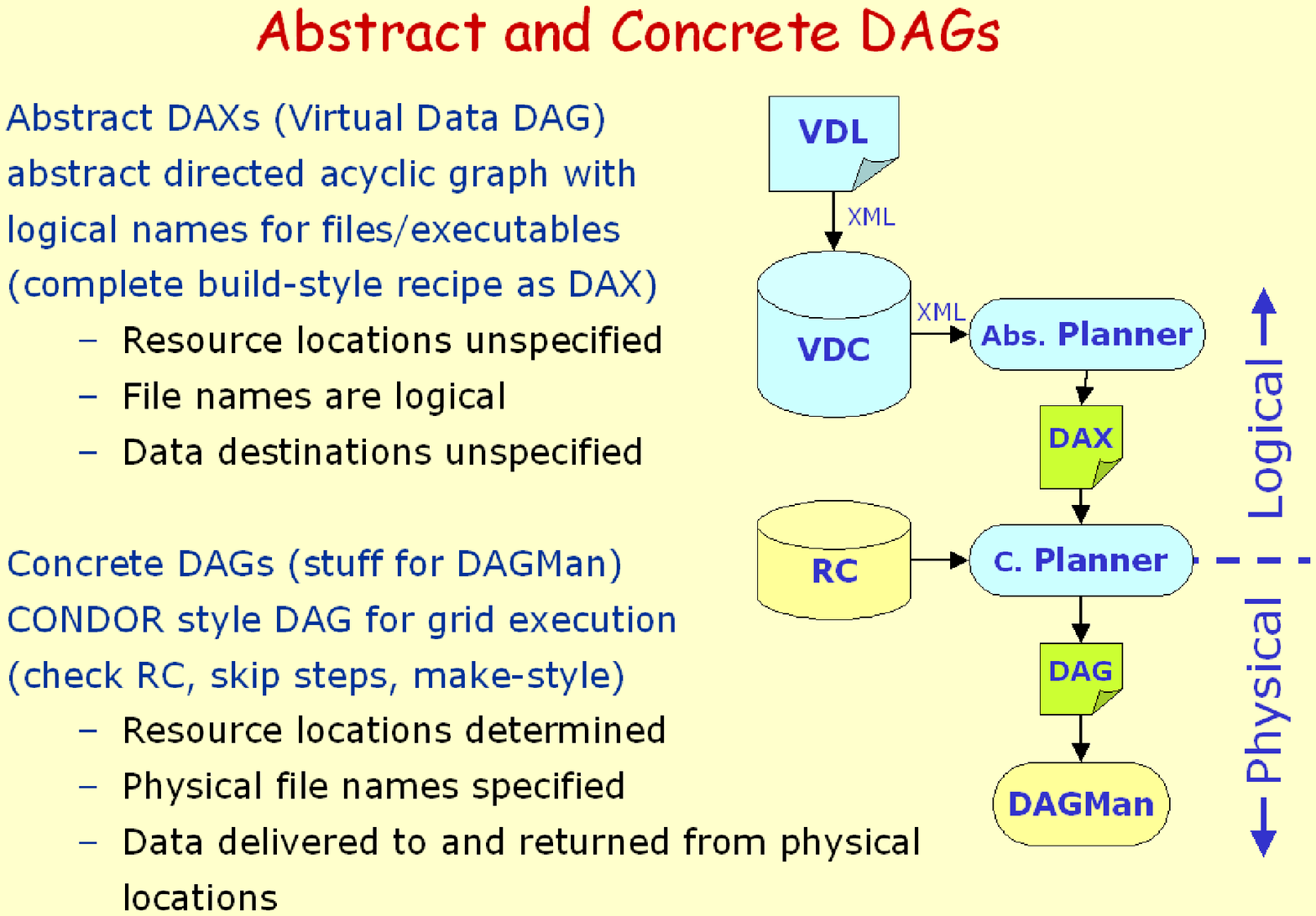}}
\caption{Steps in the derivation of a data product.} \label{dbchep03-08}
\end{figure*}
To experiment with and explore the benefits of data derivation tracking and
virtual data management, a virtual data system called
{\tt Chimera}~\cite{chimera} is under active development
in the GriPhyN project~\cite{griphyn}.
A persistent {\em virtual data catalog} (VDC), based on a relational virtual
data schema,
provides  a compact and expressive representation of the computational
procedures used to derive data, as well as invocations of those procedures
and the datasets produced by those invocations.

\begin{figure*}[t]
\centering
\resizebox{0.95\textwidth}{0.47\textheight}{\includegraphics{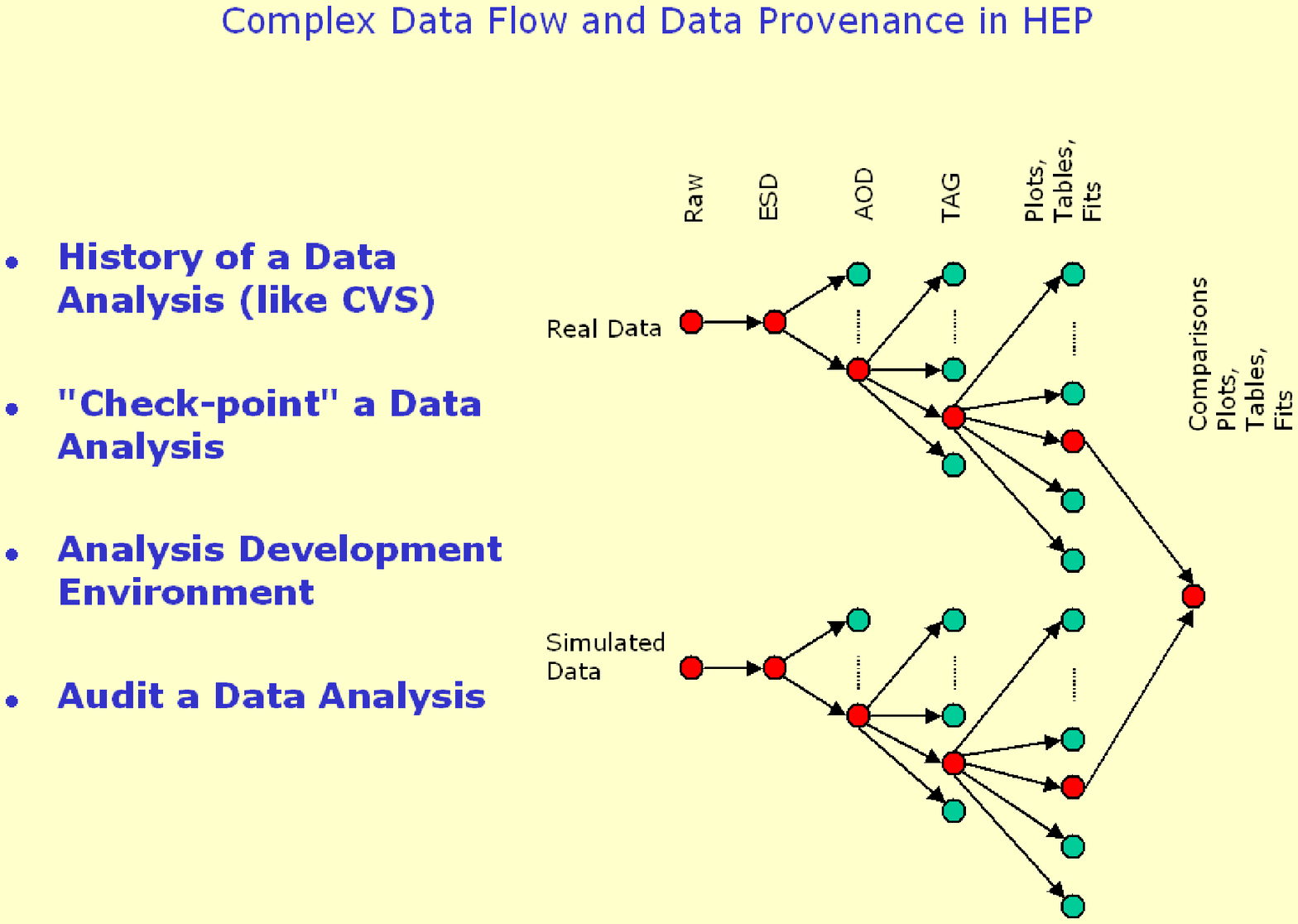}}
\caption{Example of a collaborative analysis environment.} \label{dbchep03-18}
\end{figure*}
Applications access {\tt Chimera} via a {\em virtual data language} 
(VDL), which supports both {\em data definition} statements, used for populating
a {\tt Chimera} database and for deleting and updating virtual data definitions,
and {\em query} statements, used to retrieve information from the database.
The VDL has two formats: a textual form that can be used for manual VDL
composition, and an XML form for machine-to-machine component integration.
 
Chimera VDL processing commands implement requests for constructing
and querying database entries in the VDC. These commands are implemented
in {\tt JAVA} and can be invoked from the {\tt JAVA} API or from the command
line.

The {\tt Chimera} virtual data language describes data transformation using a
function-call-like paradigm.
It defines a set of relations to
capture and formalize descriptions of how a program can be invoked, and to
record its potential and/or actual invocations. The main entities of this
language are described below:
\begin{itemize}
\item A {\em transformation} is an executable program. Associated with a
transformation is an abstract description of how the program is invoked
(e.g. executable name, location, arguments, environment). It is similar
to a ``function declaration'' in C/C++. A transformation is identified
by the tuple \mbox{[namespace]::identifier:[version \#]}.
\item A {\em derivation} represents an execution of a transformation.
It is an invocation of a transformation with specific arguments, so it is
similar to a ``function call'' in C/C++. Associated with a derivation
is the name of the corresponding transformation, the names of the data
objects to which the transformation is applied and other derivation-specific
information (e.g. values for parameters, execution time). The derivation
can be a record of how data products
came into existence {\em or} a recipe for creating them at some point in
the future.
A derivation is identified by the tuple
\mbox{[namespace]::identifier:[version range]}.
\item A {\em data object} is a named entity that may be consumed or produced
by a derivation. In the current version, a data object is a {\em logical file},
named by a {\em logical file name} (LFN). A separate replica catalog (RC) or
replica location service (RLS) is used to map from logical file names to
physical location(s) for replicas. Associated with a data object is simple
metadata information about that object.
\end{itemize}

An example of the virtual data language description for a simple pipeline with
two steps is shown in Figure~\ref{dbchep03-07}.
We define two transformations, called PYTHIA and CMSIM, which correspond to
the generation of high energy interactions with the Monte Carlo program
{\tt PYTHIA}~\cite{pitia} and the modeling of the detector response in the
CMS experiment~\cite{cms,cmssoft,cmsweb}. 
Then we define two invocations of these
transformations where the formal parameters are replaced by actual
parameters. The virtual data system detects the dependency between the output
and input files of the different steps (here file2) and automatically
produces the whole chain. This is a simple chain - the virtual data system
(VDS) has been tested
successfully on much more complex pipelines with hundreds of
derivations~\cite{sdss}.

The {\tt Chimera} system supports queries which return a representation of
the tasks to be executed as a directed acyclic graph (DAG). When executed
on a Data Grid it creates a specified data product.  The steps
of the virtual data request formulation, planning and execution process are
shown schematically in Figure~\ref{dbchep03-08}.
{\tt Chimera} is integrated with other grid services to enable the creation
of new data by executing computational schedules from database queries and
the distributed management of the resulting data.

\section{DATA ANALYSIS IN HEP}

\begin{figure*}[t]
\centering
\resizebox{0.95\textwidth}{0.45\textheight}{\includegraphics{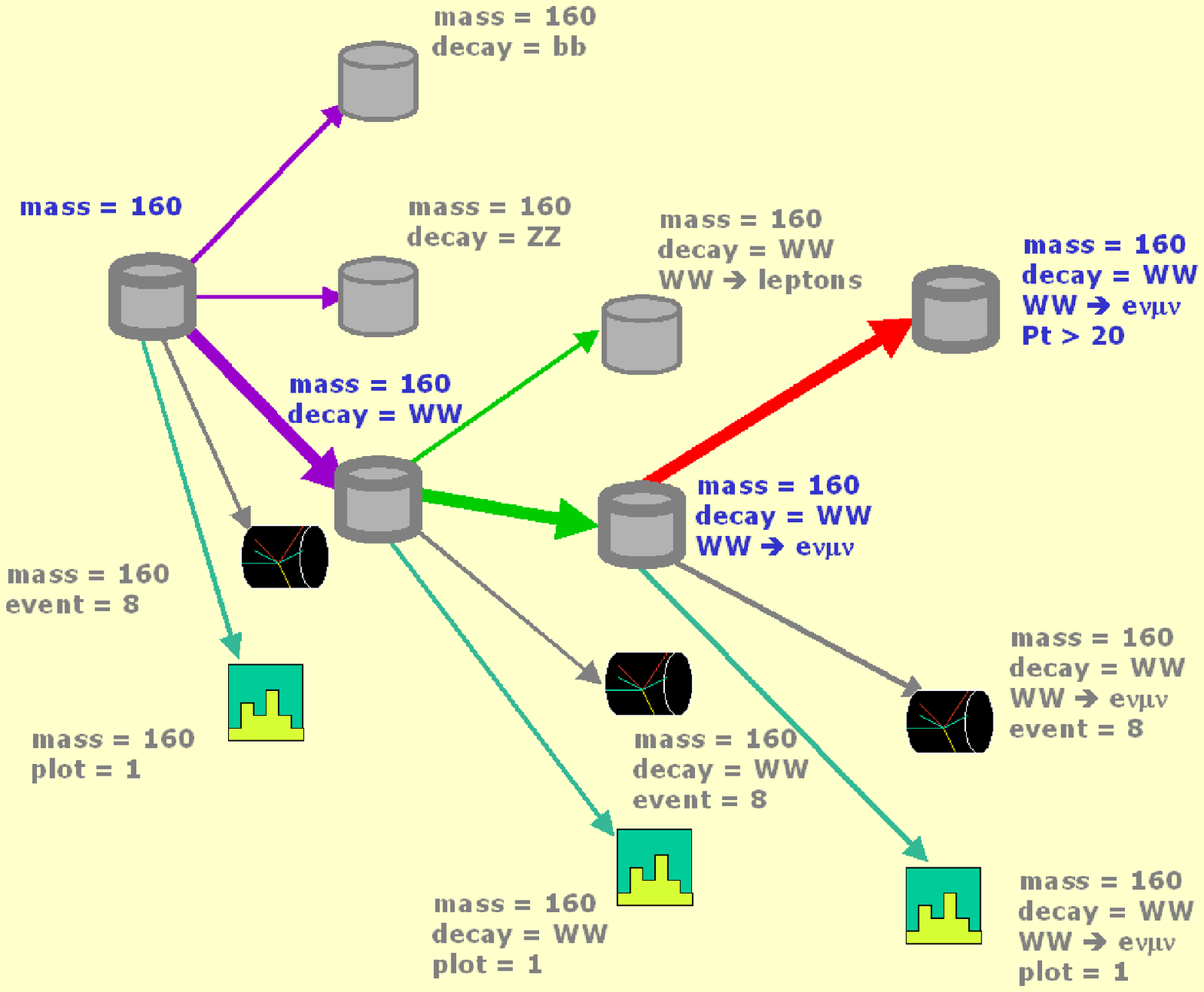}}
\caption{Example of an analysis group exploring a virtual data space.} \label{dbchep03-17}
\end{figure*}
After a high energy physics detector is triggered, the information from the
different systems is read and ultimately recorded (possibly after cleaning,
filtering and initial reconstruction) to mass storage. The high intensity
of the LHC beams usually results in more than one interaction taking
place simultaneously, so a trigger records the combined response to all particles
traversing the detector in the time window when the system is open. The first
stages in the data processing are well defined and usually tightly controlled
by the teams responsible for reconstruction, calibration, alignment,
``official'' simulation etc. The application of virtual data concepts in
this area is discussed e.g. in~\cite{cmsprod}.

In this contribution we are interested in the later stages of data processing
and analysis, when various teams and individual scientists look at the data
from many different angles - refining algorithms, updating calibrations or
trying out new approaches, selecting and analyzing a particular data set,
estimating parameters etc., and ultimately producing and publishing physics
results. Even in today's large collaborations this is a decentralized,
"chaotic" activity, and is expected to grow substantially in complexity and
scale for the LHC experiments. Clearly flexible enough systems, able to
accommodate a large user base and use cases not all of which can be foreseen
in advance, are needed. Here we explore the benefits that a virtual data
system can bring in this vast and dynamic field.

An important feature of analysis systems is the ability to build scripts
and/or executables ``on the fly'', including user supplied code and parameters.
The user should be in position to modify the inputs on her/his desk(lap)top
and request a derived data product, possibly linking with preinstalled 
libraries on the execution sites. A grid-type system can store large
volumes of data at geographically remote locations and provide the necessary
computing power for larger tasks. The results are returned to the user or
stored and published from the remote site(s). An example of this vision
is shown in Figure~\ref{dbchep03-18}. The {\tt Chimera} system can be used as
a building block for a collaborative analysis environment, providing
``virtual data logbook'' capabilities and the ability to
explore the metadata associated with different data products.

To explore the use of virtual data in HEP analysis, we take as a concrete
(but greatly simplified) example an analysis searching for the Higgs
boson at the LHC. The process begins with an analysis group that defines a
virtual data space for {\em future}
use by it's members. At the start, this space is populated solely with virtual
data definitions, and contains no materialized data products at all. A subgroup
then decides to search for Higgs candidates with mass around 160 GeV.
It selects candidates
for Higgs decaying to $W^+W^-$ and $ZZ$ bosons, $\tau^+\tau^-$ leptons and
$b\bar b$ quarks. Then it concentrates on the main decay channel 
$H \rightarrow W^+W^-$. To suppress the background, only events where both
Ws decay to leptons are selected for further processing.
Then the channel $WW \rightarrow e\nu \mu\nu$
is picked up as having low background. At each stage in the analysis 
interesting events can be visualized and plots for all quantities of
interest can be produced. Using the virtual data system all steps can be
recorded and stored in the virtual data catalog. Let us assume that a
new member joins the group. It is quite easy to discover exactly what
has been done so far for a particular decay channel, to validate how it was
done, and to refine the analysis.
A scientist wanting to dig deeper can add a new derived data branch
by, for example, applying a more sophisticated selection, and continuing to
investigate down the new branch (see Figure~\ref{dbchep03-17}).
Of course, the results of the group can be shared easily with other teams
and individuals in the collaboration, working on similar topics, providing
or re-using better algorithms etc. The starting of new subjects will
profit from the availability of the accumulated experience.
At publication time it will be much easier to perform an accurate audit of the
results, and to work with internal
referees who may require details of the analysis or additional checks.

\section{PROTOTYPES}

\begin{figure*}[t]
\centering
\resizebox{0.95\textwidth}{0.45\textheight}{\includegraphics{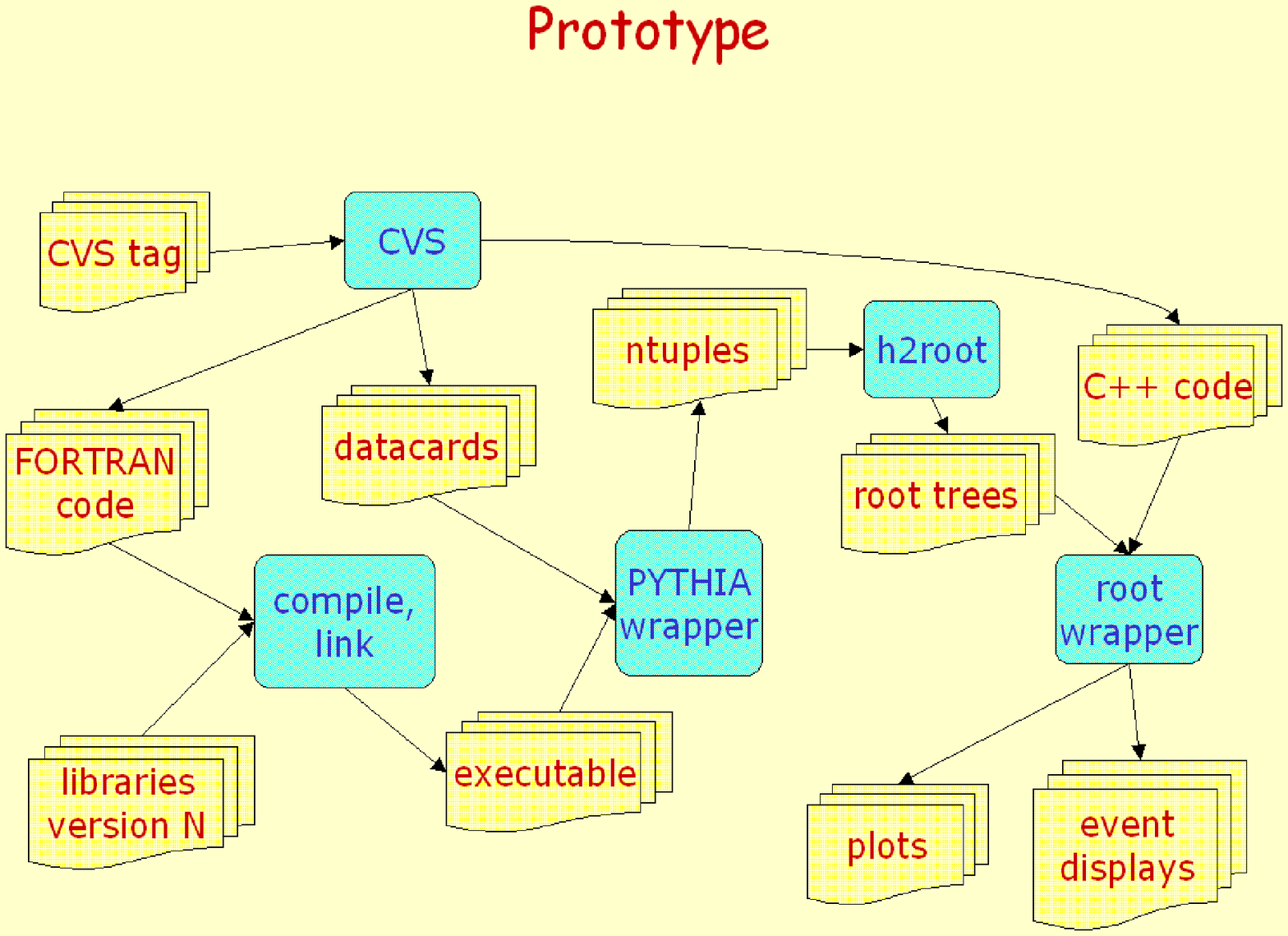}}
\caption{Analysis prototype.} \label{dbchep03-11}
\end{figure*}
In this section we describe the process of incorporating {\tt Chimera} in a
prototype of real analysis system, and examine some of the issues that arise.
In this study we use events generated with {\tt PYTHIA}
(or the CMS {\tt PYTHIA} implementation in {\tt CMKIN}), and analyze,
histogram and visualize them with the object-oriented data analysis framework
{\tt ROOT}~\cite{root}.
In the basic {\tt Chimera} implementation, the transformation is a
pre-existing program. Using the flexibility of the virtual data system,
we design our prototype with ``strong'' data provenance by using additional
steps in the pipeline. The Concurrent Version System (CVS) is well suited
to provide version control for a rapid development by a large team and to
store, by the mechanism of tagging releases, many versions so that they
can be extracted in exactly the same form even if modified, added or
deleted since that time. In our design we use wrappers (shell scripts)
at all stages in order to make the system more dynamic.
In a first step we provide a tag to the VDS and (using CVS) extract the
{\tt FORTRAN} source code and the library version number for the second,
the data cards for the third and the {\tt C++} code for the last step.
In the second step we compile and link {\tt PYTHIA} ``on the fly'', using
the library version as specified above. In the third step we generate
events with the executable and the datacards from the first two steps.
In the next, rather technical, step, we convert the generated events,
which are stored in column-wise ntuples using {\tt FORTRAN} calls to
HBOOK, to {\tt ROOT} trees for analysis. In the final step we execute a
{\tt ROOT} wrapper which takes as input the {\tt C++} code to be run
on the generated events and produces histograms or event displays.
After we define the transformations and some derivations to produce
a given data product, the {\tt Chimera} system takes care of all
dependencies, as shown in the DAG of Figure~\ref{dbchep03-11},
which is ``assembled'' and run automatically.

The Chimera configuration files describing our installed transformations are
presented below for reference.
\begin{widetext}
\begin{verbatim}
Transformation catalog (expects pre-built executables)
#pool  ltransformation physical transformation      environment String
local  hw              /bin/echo                    null
local  pythcvs         /workdir/lhc-h-6-cvs         null
local  pythlin         /workdir/lhc-h-6-link        null
local  pythgen         /workdir/lhc-h-6-run         null
local  pythtree        /workdir/h2root.sh           null
local  pythview        /workdir/root.sh             null
local  GriphynRC       /vdshome/bin/replica-catalog JAVA_HOME=/vdt/jdk1.3;VDS_HOME=/vdshome
local  globus-url-copy /vdt/bin/globus-url-copy     GLOBUS_LOCATION=/vdt;LD_LIBRARY_PATH=/vdt/lib
ufl    hw              /bin/echo                    null
ufl    GriphynRC       /vdshome/bin/replica-catalog JAVA_HOME=/vdt/jdk1.3.1_04;VDS_HOME=/vdshome
ufl    globus-url-copy /vdt/bin/globus-url-copy     GLOBUS_LOCATION=/vdt;LD_LIBRARY_PATH=/vdt/lib
\end{verbatim}
\end{widetext}

\begin{figure*}[t]
\centering
\resizebox{0.60\textwidth}{0.44\textheight}{\includegraphics{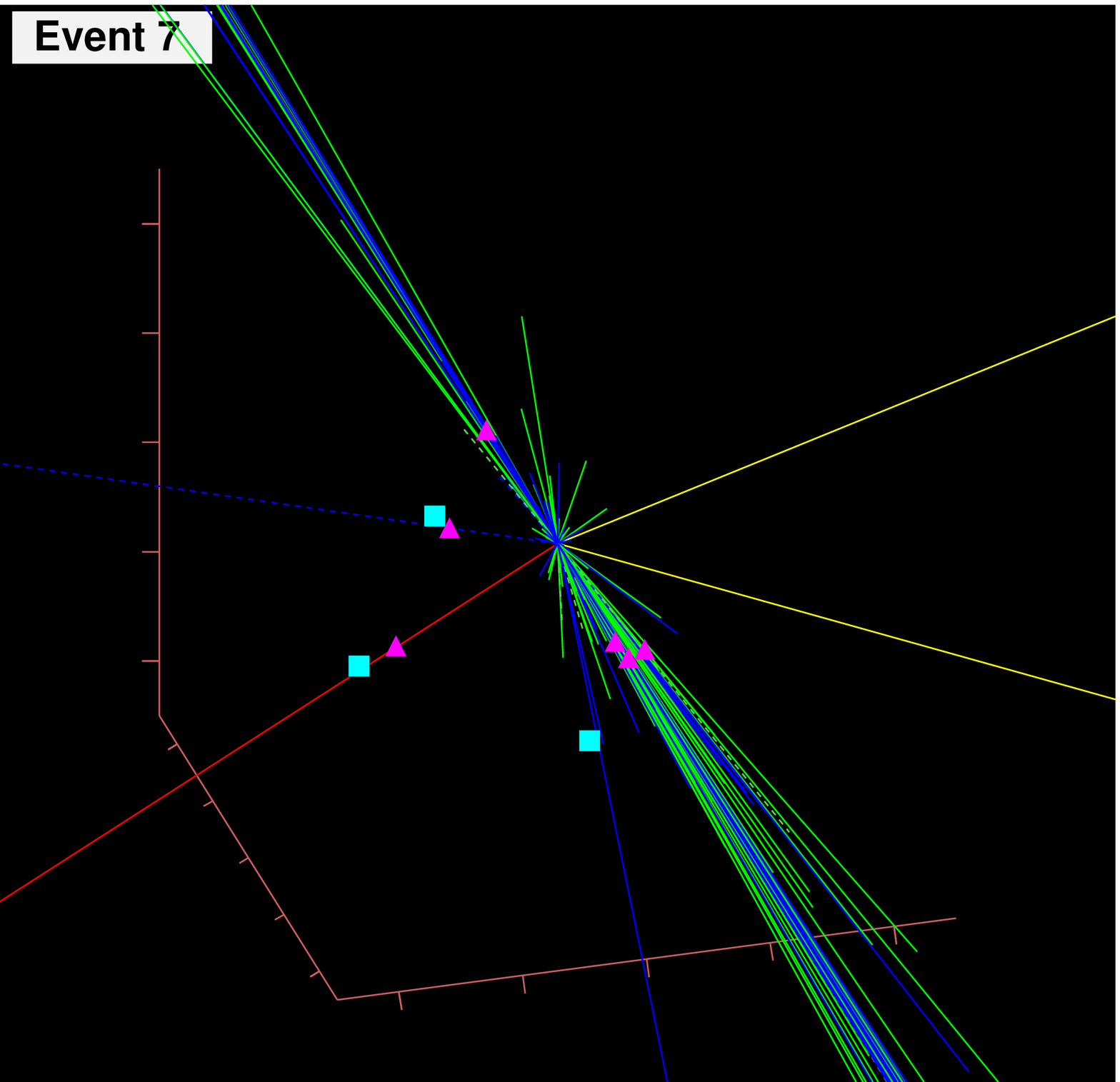}}
\caption{Event displays at LHC: $H \rightarrow W^+W^- \rightarrow e\nu\mu\nu$.
The meaning of the colors: electrons - blue dashed, photons - blue,
muons - red, neutrinos - yellow, charged hadrons - bright green,
neutral long-lived (n, $K_L$) - green dashed,
reconstructed jets: with the LUND algorithm - hot pink, with 
the cell algorithm (only detected particles) - aqua.}
\label{dbchep03-event7}
\end{figure*}
In our implementation we use MySQL as the persistent store for
the virtual data catalog.
The transformations are executed using a {\tt Chimera} tool called
the {\em shell planner}, which permits rapid prototyping by VDL processing
through execution on a local machine rather than on a full-scale grid.
Alternatively, the system can produce
a DAG which can be submitted to a local Condor~\cite{condor} pool or to a grid
scheduler.

\begin{figure*}[t]
\centering
\resizebox{0.60\textwidth}{0.44\textheight}{\includegraphics{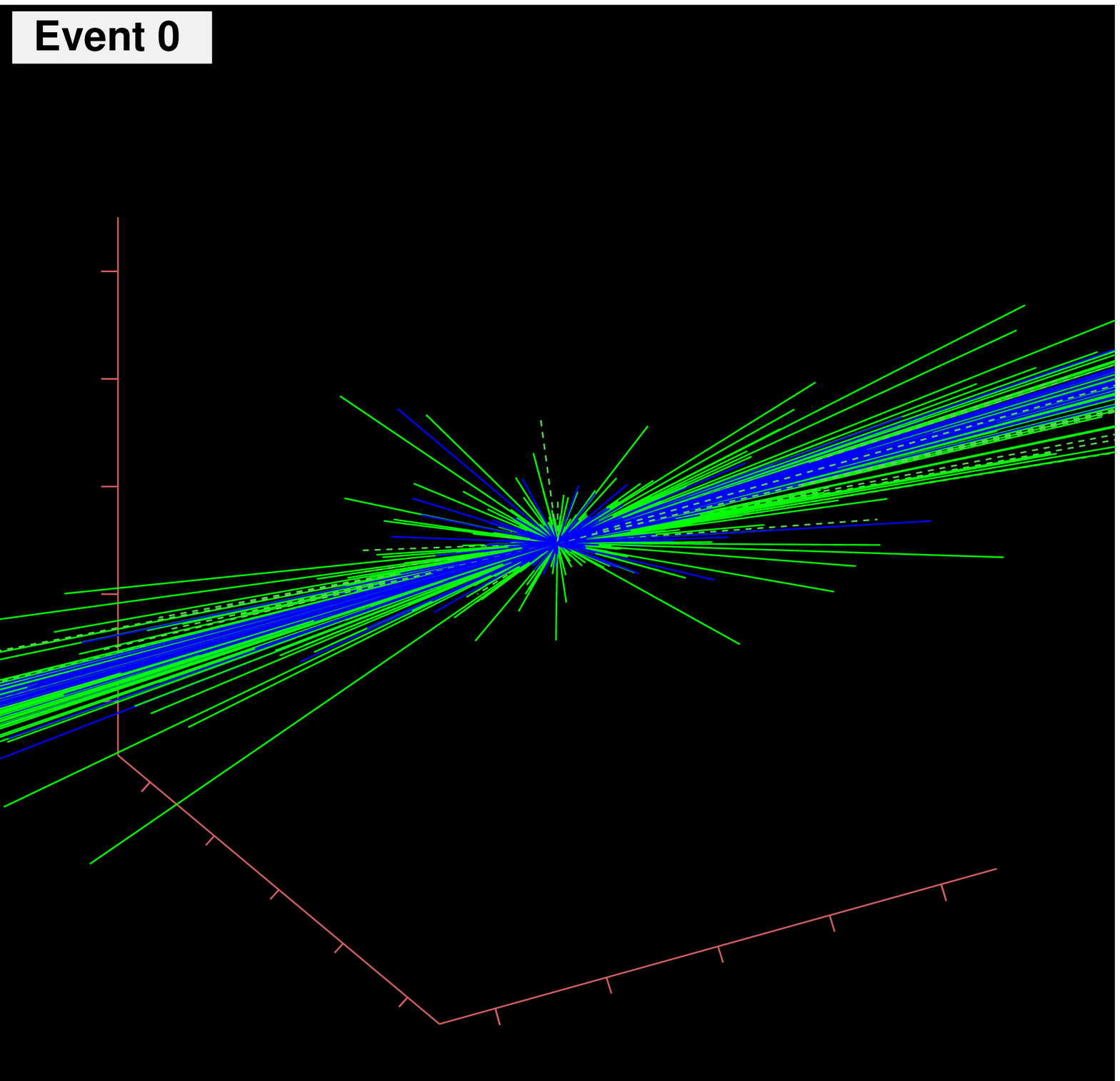}}
\caption{Event displays at LHC: a high multiplicity jet event.}
\label{dbchep03-event0}
\end{figure*}

In the last step of the illustrated derivation graph we analyze the
generated events
using the rich set of tools available in {\tt ROOT}. Besides selecting
interesting events and plotting the variables describing them, we
develop a light-weight visualization in C++ based on the {\tt ROOT}
classes. Using this tool the user can rotate the event in 3D,
produce 2D projections etc. Examples are shown in Figures~\ref{dbchep03-event7}
and~\ref{dbchep03-event0}.

\section{OUTLOOK}

We have developed a light-weight {\tt Chimera}/{\tt PYTHIA}/{\tt ROOT} prototype
for building executables "on the fly", generating 
events with {\tt PYTHIA} or {\tt CMKIN}, analyzing, plotting and visualizing
them with {\tt ROOT}. Our experience shows that {\tt Chimera} is a 
great integration tool. We are able to build a system from components using
CVS, MySQL, {\tt FORTRAN} code ({\tt PYTHIA}) and {\tt C++} code ({\tt ROOT}).
The data provenance is fully recorded and can be accessed, discovered
and reused at any time. The results reported here are a snapshot of work in
progress which is continuing to evolve both in {\tt Chimera} capabilities and
their application to CMS analysis.

This work can be extended in several directions:
\begin{itemize}
\item collaborative workflow management
\item automatic generation of derivation definitions from interactive ROOT
 sessions
\item an interactively  searchable metadata catalog of virtual data information
\item a more powerful abstractions for datasets, beyond simple files
\item control of all phases in the solving of multi-step CPU intensive
scientific problems (e.g. the study of parton density function
uncertainties~\cite{Bourilkov:2003kk})
\item integration with the CLARENS system~\cite{clarens} for remote data access
\item integration with the {\tt ROOT}/{\tt PROOF} system for parallel
analysis of large data sets
\end{itemize}

Interested readers can try out
the prototype demo, which at the time of this writing is available at the
following URL:

\begin{widetext}
\begin{verbatim}
grinhead.phys.ufl.edu/~bourilkov/pythdemo/pythchain.php .
\end{verbatim}
\end{widetext}

\begin{acknowledgments}
This work is supported in part by the United States National Science Foundation
under grants NSF ITR-0086044 (GriPhyN) and NSF PHY-0122557 (iVDGL).

\end{acknowledgments}


\end{document}